\documentclass[twocolumn,showpacs,preprintnumbers,amsmath,amssymb]{revtex4}

\usepackage{graphicx}

\date{\today}

\begin{document}

\title{Quantum-``classical'' correspondence in a nonadiabatic-transition system}

\author{Hiroshi Fujisaki}\email{fujisaki@bu.edu}
\affiliation{%
Department of chemistry,
Boston University, 590 Commonwealth Ave.,
Boston, 02215, Massachusetts, USA
}%

\begin{abstract}

A nonadiabatic-transition system which exhibits ``quantum chaotic'' behavior 
[Phys.~Rev.~E {\bf 63}, 066221 (2001)]
is investigated from quasi-classical aspects.
Since such a system does not have a naive classical limit,
we take the mapping approach by Stock and Thoss 
[Phys.~Rev.~Lett.~{\bf 78}, 578 (1997)] to represent the quasi-classical
dynamics of the system.
We numerically show that 
there is a sound 
correspondence between the quantum chaos and classical chaos for the system.

\end{abstract}


\maketitle


Nonadiabatic transition (NT)
is a very fundamental concept in physics and chemistry \cite{Nakamura02,TAWM00}.
In atomic, molecular, and  chemical physics literature,
NTs occur as a breakdown of the 
Born-Oppenheimer (BO) approximation, 
which is essentially an adiabatic approximation to 
solve quantum systems with many degrees of freedom.
It is still a tough problem
to analyze the properties of NTs
especially for multidimensional systems.  
One reason preventing us from deeper understanding of 
NT is the lack of a naive quantum-classical correspondence for NT
like tunneling phenomena \cite{UT96,SI98}.
One way to get the classical picture for a NT system
is to go back to the original system before the BO approximation:
Fuchigami and Someda investigated dynamical 
properties of H$^+_2$ from classical points of view 
by treating an electron and nuclei as dynamical variables~\cite{FS03}.
Though there is a full quantum study for such a small system to compare with \cite{KKF99},
this ``purist'' way cannot be easily applied to much more ``complex'' systems.

The mapping method recently advocated by Stock and Thoss \cite{ST97}
can circumvent this deficiency.
(This is reminiscent of the Meyer-Miller method \cite{MM79}.)
Their method is as follows: After the BO approximation, 
the discrete electronic degrees of freedom are mapped 
onto the Schwinger bosons \cite{JJSakurai}. 
Since all the degrees of freedom 
become just bosons, the total system is rather easily treated 
semiclassically or quasi-classically.
Using this method semiclassically, one can obtain, e.g., 
absorption spectra even for a pyrazin molecule with
24 degrees of freedom \cite{TMS00}. 
One can use it quasi-classically by solving the equations of 
motion derived from a mapping Hamiltonian. 
This is a very easy way to simulate NT systems because the 
additional number of degrees of freedom for electronic parts is rather small.
Using the periodic orbit theory \cite{Gutzwiller90} or 
the adiabatic switching method \cite{Reinhardt90}, 
one can obtain even quantum eigenenergies and eigenstates, 
in principle \cite{ZGE86}.

On the other hand, multidimensional NT systems like 
Jahn-Teller molecules \cite{KDC84} are known 
to show ``quantum chaotic'' behavior \cite{Gutzwiller90}. 
Fujisaki and Takatsuka investigated this problem deeply
employing the two-mode-two-state (TMTS) system which is  
considered as a system with two vibrational modes and two 
electronic states \cite{FT01b}.
They calculated the statistical properties of the 
eigenenergies and eigenfunction for the TMTS system,
and found that the system becomes strongly ``quantum chaotic'' under 
a certain condition. 
In addition, they showed that the chaos is not just a reflection of the lower adiabatic system
nor that of the diabatic systems.
(On the other hand, the chaos of all previous studies is just a reflection 
of the lower adiabatic systems \cite{KDC84}.)
This means that conventional classical descriptions do not help to explain the quantum 
chaotic behavior.
Hence this system deserves to be further studied from the 
``mapping'' (extended classical) points of view.
Though there are some studies which investigated chaotic properties 
of this kind of mixed quantum-classical systems \cite{BE94}, 
our focus here is a quantum-``classical'' correspondence (if any)
for the TMTS system.

\begin{figure}[htbp]
\begin{center}
\includegraphics[scale=0.45]{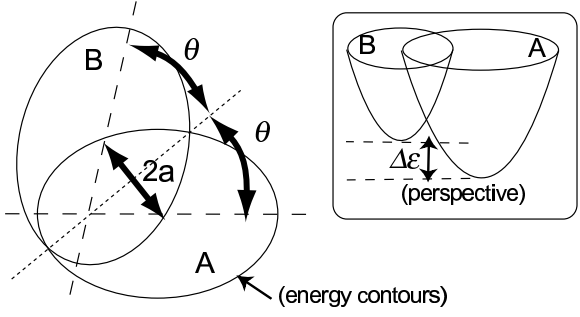}
\caption{
\label{fig:TMTS}
A shematic representation of the TMTS system. The distance between the minima of the 
potential is $2 a$, and the angle between the relevant crossing seam (dotted line) 
and the primary axis of each potential (dashed line) is $\theta$.
Inset: The perspective of the TMTS system.
The potential minima are different with 
$\Delta \epsilon = \epsilon_B-\epsilon_A=0.173$.
}
\end{center}
\end{figure}

The TMTS system \cite{FT01b} first introduced by Heller \cite{Heller90}
is described by the following Hamiltonian:
\begin{equation}
{\cal H}_{\rm TMTS}=\left( 
\begin{array}{cc}
T_{{\rm kin}}+V_{A} & J \\ 
J & T_{{\rm kin}}+V_{B}
\end{array}
\right),  
\label{chaos_Heller}
\end{equation}
where $T_{\rm kin}$ is the kinetic energy,
$V_i$ ($i=A,B$) is the potential energy for state $i$
defined by
\begin{eqnarray}
T_{{\rm kin}} 
&=& \frac{1}{2}(p_{x}^{2}+p_{y}^{2}),  
\\
V_A 
&=& 
\frac{1}{2}(\omega _{x}^{2} x^{2}+\omega _{y}^{2} y^{2})+\epsilon _A,
\\
V_B 
&=& 
\frac{1}{2}(\omega _{x}^{2} \xi^{2}+\omega _{y}^{2} \eta^{2})+\epsilon _B,
\end{eqnarray}
with
\begin{eqnarray}
\xi 
&=& 
(x +2 a \sin \theta) \cos 2\theta +(y-2 a \cos \theta) \sin 2 \theta,  
\\
\eta &=& 
-(x +2 a \sin \theta) \sin 2\theta +(y-2 a \cos \theta) \cos 2 \theta.  
\label{eq:xieta}
\end{eqnarray}
Note that we have just used a harmonic potential for each state.
For the geometrical meaning of the parameters, see Fig.~\ref{fig:TMTS}.
Here the Duschinsky angle $\theta$ \cite{TLL03} 
and the nonadiabatic coupling constant $J$ are two 
important parameters for the system.
If these are appropriately chosen, the system becomes strongly ``quantum chaotic'',
i.e., the nearest-neighbor energy-level spacing distribution becomes 
the Wigner type and the amplitude distribution of the eigenstates 
becomes Gaussian \cite{FT01b}.

The mapping Hamiltonian \cite{ST97,TMS00} 
for this system is
\begin{eqnarray}
H_{\rm map}
= 
T_{\rm kin}
+ N_A V_A +N_B V_B 
+J (x_A x_B+p_A p_B)
\end{eqnarray}
with
\begin{eqnarray}
N_i
= \frac{x_i^2+p_i^2}{2}-\gamma
\quad 
(i=A,B).
\end{eqnarray}
Note that $N_A+N_B=1$. From this relation, one might consider 
$N_i$ ($i=A,B$) 
as a probability but it is not the case. This is because 
the numerical range for $N_i$ is $-\gamma < N_i < 1 + \gamma$. 
The equations of motion for this Hamiltonian are derived 
as follows:
\begin{eqnarray}
\frac{d}{dt} x_A 
&=& V_A p_A + J p_B, \quad 
\frac{d}{dt} p_A 
= -V_A x_A - J x_B,
\\
\frac{d}{dt} x_B 
&=& V_B p_B + J p_A, \quad 
\frac{d}{dt} p_B 
= -V_B x_B - J x_A,
\\
\frac{d}{dt}x &=& p_x,
\quad
\frac{d}{dt}p_x= -\frac{\partial V_A}{\partial x} 
N_A-\frac{\partial V_B}{\partial x}N_B,
\\
\frac{d}{dt}y &=& p_y,
\quad
\frac{d}{dt}p_y= -\frac{\partial V_A}{\partial y} 
N_A-\frac{\partial V_B}{\partial y} N_B.
\end{eqnarray}

As a reference system, 
we take the lower adiabatic system as in \cite{FT01b}.
The lower adiabatic systems is defined by the following 
Hamiltonian:
\begin{eqnarray}
H^{-}_{\rm ad} 
= T_{\rm kin} +V_{\rm ad}^-,
\end{eqnarray}
with
\begin{eqnarray}
V_{\rm ad}^-
= 
\frac{V_A+V_B}{2}-
\sqrt{ \left( \frac{V_A-V_B}{2} \right)^2 +J^2}
\end{eqnarray}

Since the TMTS system can be quantum chaotic as mentioned above, 
we investigate the chaotic properties for the mapping system
and the lower adiabatic system.
As an indicater of chaos, it is very natural to take Lyapunov 
exponents \cite{Ott02}.
We focus on a finite-time maximum Lyapunov exponent 
calculated as 
\begin{equation}
\lambda_{\rm max}(T) \simeq  
\frac{1}{T} \log \frac{\Delta d(T)}{\Delta d(0)}
\end{equation}
with
\begin{eqnarray}
\Delta d(t)^2
&=& 
\Delta \tilde{x}_A(t)^2 + \Delta \tilde{p}_A(t)^2 
+ \Delta \tilde{x}_B(t)^2 + \Delta \tilde{p}_B(t)^2
\nonumber
\\
&+&
 \Delta \tilde{x}(t)^2 + \Delta \tilde{p}_x(t)^2 
+ \Delta \tilde{y}(t)^2 + \Delta \tilde{p}_y(t)^2
\end{eqnarray}
where $
\tilde{x}_A =x_A/\sqrt{2+2 \gamma}, \quad
\tilde{p}_A =p_A/\sqrt{2+2 \gamma}, \quad
\tilde{x}_B =x_A/\sqrt{2 \gamma}, \quad
\tilde{p}_B =p_A/\sqrt{2 \gamma}, \quad
\tilde{x}= \omega_x x/\sqrt{2 E}, \quad
\tilde{y}= \omega_y y/\sqrt{2 E}, \quad
\tilde{p}_x= p_x/\sqrt{2 E}, \quad
\tilde{p}_y= p_y/\sqrt{2 E},
$
and $E$ is the total energy for the system.
Here tilde variables are introduced for normalization 
and $\Delta$ means a distance between a trajectory and 
its auxiliary one.
In the following we take the typical value of $\gamma$, 
i.e., $\gamma=1/2$ \cite{ST97}.
For the numerical calculation of Lyapunov exponents, 
the method by Benettin {\it et al}. \cite{BGS76} is employed,
i.e., we calculate $\lambda_{\rm max}(T)$ for a finite $T$, 
then shorten the distance $\Delta d(T)$ to $\Delta d(0)$,
and run the trajectory again and so on. 
In this study, we took $T=24$ because of the 
numerical divergence of the $\Delta d(T)$ for larger 
values of $T$.

\begin{figure}[htbp]
\hfill
\hfill
\begin{center}
\includegraphics[scale=0.6]{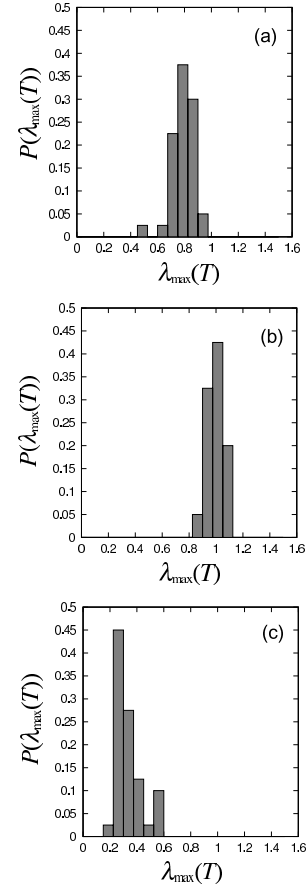}
\caption{
\label{fig:dist_TMTS}
$J$ dependence of the distribution of the Lyapunov exponents
for the mapping system. 
The nonadiabatic coupling is 
(a) $J=0.3$, (b) $J=1.5$, and (c) $J=7.5$.
The finite time is $T=24$, and 
the iterative number of the time average is 10.
The Duschinsky angle is $\theta=\pi/3$.
}
\end{center}
\end{figure}
\begin{figure}[htbp]
\begin{center}
\includegraphics[scale=0.6]{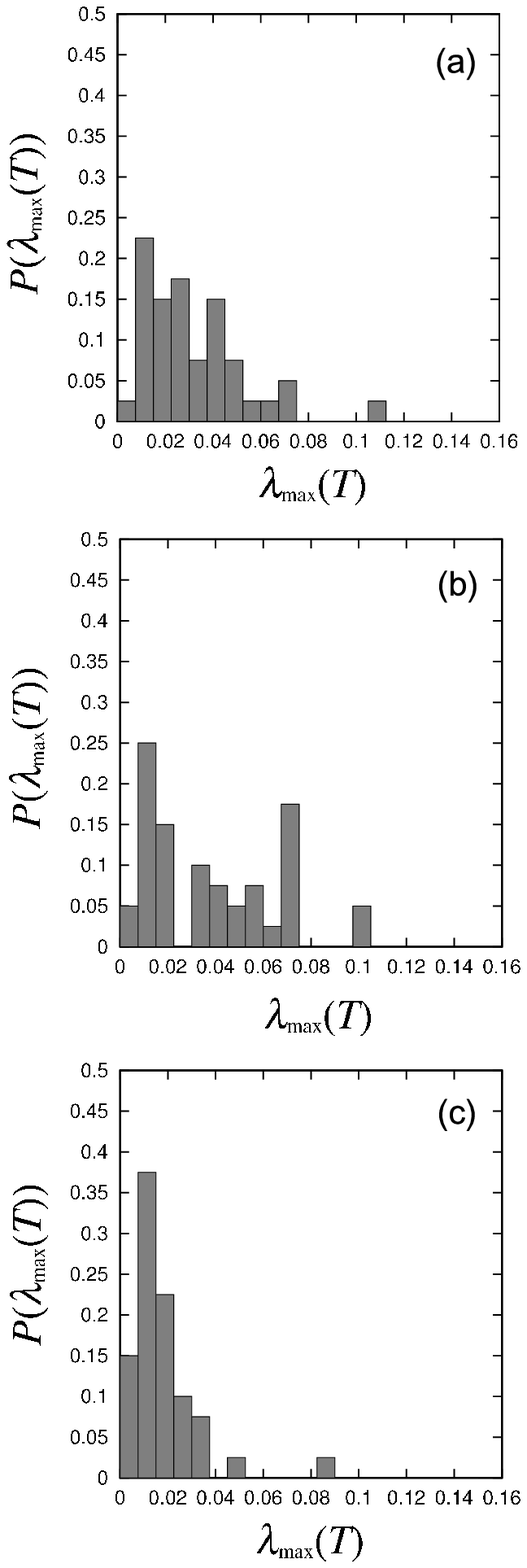}
\caption{
\label{fig:dist_lad}
$J$ dependence of the distribution of the Lyapunov exponents
for the lower adiabatic system.
The nonadiabatic coupling is 
(a) $J=0.3$, (b) $J=1.5$, and (c) $J=7.5$.
The finite time is $T=24$, and 
the iterative number of the time average is 10.
The Duschinsky angle is $\theta=\pi/3$.
}
\end{center}
\end{figure}

\begin{figure}[htbp]
\begin{center}
\includegraphics[scale=0.6]{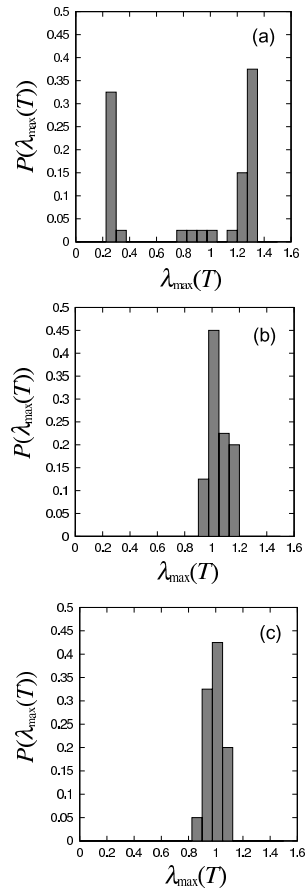}
\caption{
\label{fig:dist_theta}
$\theta$ dependence of the phase space averaged Lyapunov exponent
for the mapping system with $J=1.5$.
The Duschinsky angle is 
(a) $\theta=0.0$, (b) $\theta=\pi/6$, and (c) $\theta=\pi/3$.
The finite time is $T=24$, and 
the iterative number of the time average is 10.
}
\end{center}
\end{figure}

Following the previous studies \cite{FT01b}, 
we concentrate on a rather high energy region around $E=E_0=28$.
(This energy is much higher than that around the crossing seam region.)
Varying the Duschinsky angle $\theta$ and the nonadiabatic coupling
constant $J$, we calculate the distribution of the finite-time maximum 
Lyapunov exponent $\lambda_{\rm max}(T)$. For simplicity, hereafter, 
we call $\lambda_{\rm max}(T)$ just as a Lyapunov exponent.
Since the phase space has a structure especially for the 
lower adiabatic system, globally averaged Lyapunov exponents 
are not so useful. Instead, we investigate the 
properties of the distribution of the Lyapunov exponents, 
which reflects phase space structure of the system.
(Remember that
the Berry-Robnik distribution, which reflects 
the phase space volume of chaotic seas,
is useful for mixed systems \cite{MHA01}.) 
To this end, we prepare an initial ensemble of particles
which are sampled from a part of the equi-energy potential surface 
$V_A(x,y)=E_0$ with constraints $p_x=p_y=0$ and $y< x \tan \theta +a/\cos \theta$.
(The latter constraint means that we only take points below the crossing 
seam. See Fig.~\ref{fig:TMTS}.)
We take 40 sample points from this curve and calculate the histogram for the 
Lyapunov exponents.
We believe that these sample points represent 
a typical situation of the TMTS system because, at least, 
the characteristic of the lower adiabatic system can be 
understood from these sampling points \cite{FT01b}.

First we fix $\theta = \pi/3$, 
and investigate the $J$ dependence 
of the chaotic properties.
Let us summarize the corresponding quantum results:
the nearest-neighbor (energy-level) spacing distribution 
is Wigner for $J=1.5$ (3rd row of Fig.~3 in \cite{FT01b}),
whereas it is rather Poisson for $J=7.5$ (3rd row of Fig.~4 in \cite{FT01b}),
and rather mixed for $J=0.3$ (3rd row of Fig.~2 in \cite{FT01b}).
Figure \ref{fig:dist_TMTS} shows the distribution of the Lyapunov 
exponents for the TMTS system.
The distribution for $J=1.5$ has a sharp peak around 1, 
whereas that for $J=7.5$ has a rather broad peak around 0.4,
and that for $J=0.3$ is also a little bit broad.
This corresponds to the quantum results, at least, qualitatively. 
On the other hand, 
Fig.~\ref{fig:dist_lad} shows the distribution of the Lyapunov 
exponents for the lower adiabatic system.
As one can see, the values themselves are much smaller than 
those for the TMTS system and it is difficult 
to distinguish three distributions. 
This also corresponds to the quantum mechanical calculation 
for the lower adiabatic system (Fig.~9 in \cite{FT01b}).
From this comparison, it is reasonable to conclude 
that there is a sound quantum-``classical'' correspondence
between the TMTS system and the mapped system in view 
of their ``chaotic'' properties.

Next we investigate the $\theta$ dependence of the 
chaotic properties while fixing $J=1.5$.
Although there is a strong peak around $\lambda_{\rm max}(T) \simeq 1.3$ 
as shown in Fig.~\ref{fig:dist_theta} (a), 
we can see that the system with $\theta=0$ is not 
{\it globally} chaotic. This is because $x$ does not effectively coupled to $y$ 
when $\theta=0$, and the motion along $x$ axis is regular.
(On the other hand, even with $\theta=0$, the motion along $y$ axis 
can be chaotic as shown by the peak of the distribution around 1.3.)
In such a case, we do not expect that the Wigner type distribution arises 
in the corresponding quantum system,
and this is the case for the TMTS system 
(1st row of Fig.~3 in \cite{FT01b}). 
On the other hand, for intermediate Duschinsky angles 
($\theta=\pi/6, \pi/3$), the Lyapunov exponent distributions show
that the system is globally chaotic [Fig.~\ref{fig:dist_theta} (b),(c)], 
and the corresponding quantum system can have the Wiger type distribution,  
which is also confirmed numerically (2nd and 3rd rows of Fig.~3 in \cite{FT01b}). 
Of course, we have to admit that this correspondence is loosely stated,
and there remains a difficult question exactly when the quantum chaos 
begins in the parameter space. Such an issue must be addressed utilizing
semiclassical methods \cite{ST97,Gutzwiller90} 
or the adiabatic switching method \cite{Reinhardt90}.
It is also interesting to investigate phase space structure 
of this mapping system and to relate it to the nearest-neighbor 
spacing distribution 
via the Berry-Robnik distribution \cite{MHA01}.
We also hope that this study will cast a light on  
the relation between the statistical reaction theory 
for NT systems and Lyapunov spectra for them.

In this paper, 
employing the mapping approach by Stock and Thoss,
we investigated 
a nonadiabatic-transition system which exhibits ``quantum chaotic'' behavior 
from quasi-classical aspects.
By comparing the statistical properties of the quantum system with 
the Lyapunov exponent distributions of the mapping system,
we found that 
there is a sound quantum-``classical'' correspondence in the system.

The author thanks T.~Takami for comments.

\end{document}